\documentclass[%
superscriptaddress,
twocolumn,
 amsmath,amssymb,
 aps,
 prl,
floatfix,
]{revtex4-2}

\usepackage{graphicx}% Include figure files
\usepackage{dcolumn}% Align table columns on decimal point
\usepackage{bm}% bold math
\begin{document}

\graphicspath{ {./figures/} }
\preprint{APS/123-QED}

%\title{Spinons and damped phonons in spin-1/2 quantum-liquid Ba$_{4}$Ir${}_3$O${}_{10}$ Raman scattering}%

\title{Spinons and damped phonons in spin-1/2 quantum-liquid Ba$_{4}$Ir${}_3$O${}_{10}$ observed by Raman scattering.}%

%\title{Four-spinon hump and damped phonons via Raman in spin-1/2 quantum-liquid Ba$_{4}$Ir${}_3$O${}_{10}$}%
%\title{Spinons and damped phonons in spin-1/2 quantum-liquid Ba$_{4}$Ir${}_3$O${}_{10}$ observed by Raman scattering.}%
%\title{Spinons and damped phonons in the spin-orbit insulator Ba$_{4}$Ir${}_3$O${}_{10}$ via Raman}%

%Damping of phonons by magnetic fluctuations in the magnetic spin-orbit insulator

\author{Aaron Sokolik}
\affiliation{Department of Physics, University of Colorado Boulder, Boulder, CO 80309}
\author{Sami Hakani}
\affiliation{School of Physics, Georgia Institute of Technology, Atlanta, GA 30332}
\author{Susmita Roy}
\affiliation{Department of Physics, University of Colorado Boulder, Boulder, CO 80309}
\author{Nicholas Pellatz}
\affiliation{Department of Physics, University of Colorado Boulder, Boulder, CO 80309}
\author{Hengdi Zhao}
\affiliation{Department of Physics, University of Colorado Boulder, Boulder, CO 80309}
\author{Gang Cao}
\affiliation{Department of Physics, University of Colorado Boulder, Boulder, CO 80309}
\author{Itamar Kimchi}
\affiliation{School of Physics, Georgia Institute of Technology, Atlanta, GA 30332}
\author{Dmitry Reznik}
\affiliation{Department of Physics, University of Colorado Boulder, Boulder, CO 80309}

\date{October 28, 2020}% It is always \today, today,

\begin{abstract}
In spin-1/2 Mott insulators, non-magnetic quantum liquid phases are often argued to arise when the system shows no magnetic ordering, but identifying positive signatures of these phases or related spinon quasiparticles can be elusive. Here we use Raman scattering to provide three signatures for spinons in  a possible spin-orbit quantum liquid material Ba${}_4$Ir${}_3$O${}_{10}$: (1) A broad hump, which we show can arise from Luttinger Liquid spinons in Raman with parallel photon polarizations normal to 1D chains;  (2) Strong phonon damping from phonon-spin coupling via the spin-orbit interaction; and (3) the absence of (1) and (2) in the magnetically ordered phase that is produced when 2\% of Ba is substituted by Sr ((Ba${}_{0.98}$Sr${}_{0.02}$)${}_4$Ir${}_3$O${}_{10}$). The phonon damping via itinerant spinons seen in this quantum-liquid insulator suggests a new mechanism for enhancing thermoelectricity in strongly correlated conductors, through a neutral quantum liquid that need not affect electronic transport. 
\end{abstract}


\maketitle

%\tableofcontents
%%%% BEGIN PASTE

Spin-1/2 magnetic insulators can avoid the conventional fate of magnetic order, by instead forming one of various types of exotic quantum liquid states. Such quantum liquids in 2D spin systems come in variously many flavors: gapped spin liquids such as the toric code \cite{wen1991, KITAEV2006, savary2016, Zhou2017, balents2010}, gapless spin liquids such as a spinon Fermi surface \cite{savary2016, Zhou2017, balents2010}, sliding Luttinger liquids \cite{mukhopadhyay2001}, and yet other examples whether recently discovered or yet to be discovered \cite{lake2021, savary2016, Zhou2017, balents2010}. Their common features include a high degree of quantum entanglement, and (typically) exotic ``spinon''-type excitation. But beyond mere lack-of-ordering, positive signatures that identify a particular material as a quantum liquid state are difficult to access. 
Theoretical treatments of the strongly spin-orbit coupled honeycomb iridates and, more recently, $\alpha$-RuCl$_3$ have inspired a large body of experimental work that anticipates a spin liquid. \cite{KITAEV2006, chaloupka2010, Jackeli2009, witczak2014, rau2016, cao2018, hermanns2018, nasu2016,  hidenori2019}  However, there has been no clear-cut material realization of a quantum spin liquid thus far, suggesting that a different material approach may also be fruitful.

\begin{figure}[htp]
\begin{center}
\includegraphics[width=0.5\textwidth]{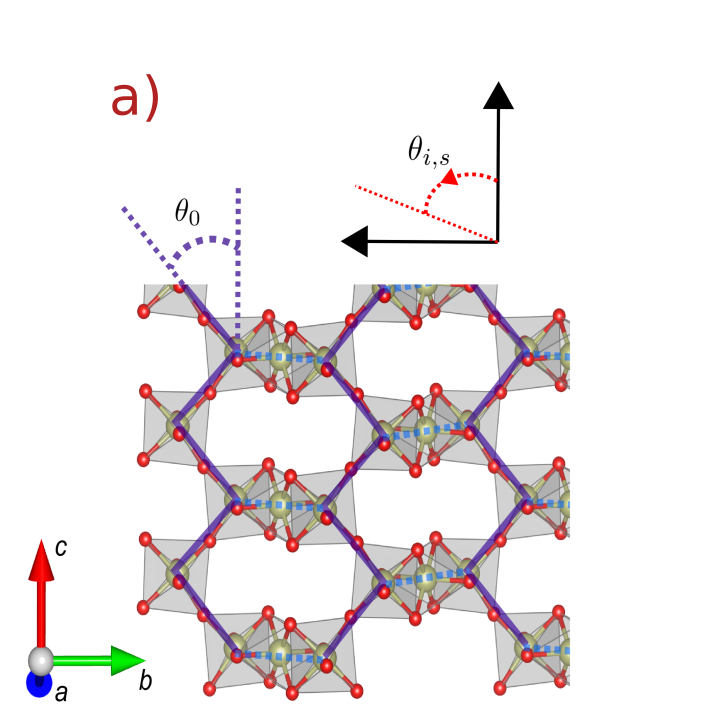}
\includegraphics[width=0.5\textwidth]{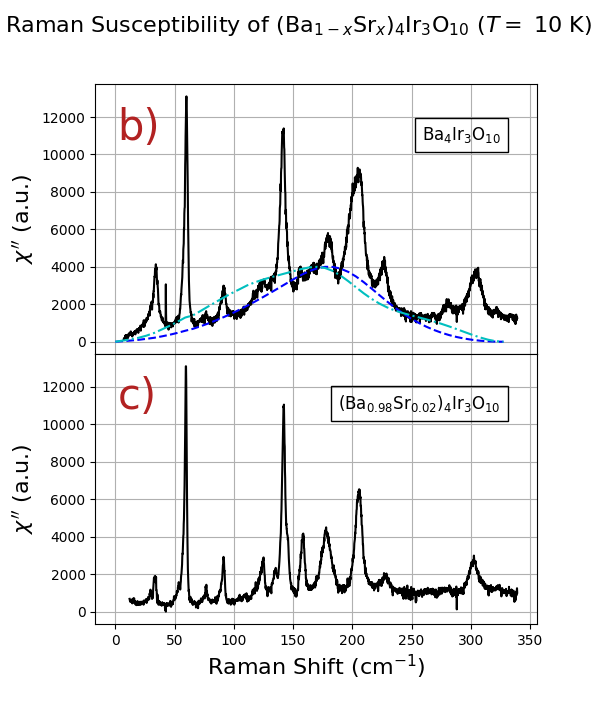}
\caption{Raman spectra of quantum liquid Ba${}_4$Ir${}_3$O${}_{10}$, and the Sr-doped magnetically ordered sister compound ((Ba${}_{0.98}$Sr${}_{0.02}$)${}_4$Ir${}_3$O${}_{10}$). (a) Ba${}_4$Ir${}_3$O${}_{10}$ is a  2D layered spin-1/2 magnet, featuring 1D zigzag chains coupled via trimers. (b,c) show Raman scattering spectra in $bb$ photon polarization.  The pristine compound features broadened phonon peaks on top of a broad hump, which is captured well by a four-spinon continuum in quantum liquid mean field theories with $J_\text{eff}/k_B = 75$ K ($R_{1}$ dashed, blue; $R_2$ dot-dashed, cyan). Non-magnetic Sr substitution in (Ba${}_{0.98}$Sr${}_{0.02}$)${}_4$Ir${}_3$O${}_{10}$ produces non-frustrated magnetic order and eliminates the phonon broadening and the 4-spinon hump.}
\label{fig:ramantemp}
\end{center}
\end{figure}

Here we report on our observation and analysis of signatures of fractionalized spinons, through low temperature Raman spectroscopy of a recently discovered \cite{cao2020} quasi-2D spin-1/2  material, Ba${}_4$Ir${}_3$O${}_{10}$. 
Ba${}_4$Ir${}_3$O${}_{10}$ adopts a monoclinic structure with a P$2_1$/c space group, which consists of Ir${}_3$O${}_{12}$ trimers (with Ir$4^+$(5d${}^5$) ions) of face-sharing IrO${}_6$ octahedra that are vertex-linked to other trimers, forming 2D sheets of a distorted square lattice in the $bc$ plane, that are stacked along the a-axis with no connectivity between the sheets (Fig.\ \ref{fig:ramantemp}a). Electrical resistivity shows a clear insulating state across the entire temperature range measured up to 400 K. %The Ba${}_4$Ir${}_3$O${}_{10}$ has been seen as a candidate for a new kind of frustrated quantum liquid arising in the magnetic insulator.
The strong spin-orbit-coupling and frustration of the Ir spin-1/2 moments conspire to form an unusual kind of highly frustrated quantum liquid state: Ba${}_4$Ir${}_3$O${}_{10}$ shows no magnetic order down to 0.2 K, despite strong antiferromagnetic interactions with Curie-Weiss temperatures ranging from 100--700 K \cite{cao2020}. 
This quantum liquid state  shows a sizable linear heat capacity with a constant offset at very low temperatures, $C\sim T + C_0$ implying that the absence of magnetic ordering is accompanied by gapless quantum-fluctuating excitations. Linear (and surprisingly small) thermal conductivity is also observed at low temperatures, again consistent with the quantum liquid state. Finally, this quantum liquid state is evidently quite fragile and disappears upon slight perturbations to the crystal: experimentally, a mere 2\% Sr substitution for Ba ((Ba${}_{0.98}$Sr${}_{0.02}$)${}_4$Ir${}_3$O${}_{10}$) precipitates conventional AFM order below 130 K,  thereby eliminating the linear-$T$ heat capacity and drastically changing magnetization and thermal conductivity.

\begin{figure}[htp]
%\begin{center}
\includegraphics[width=0.5\textwidth]{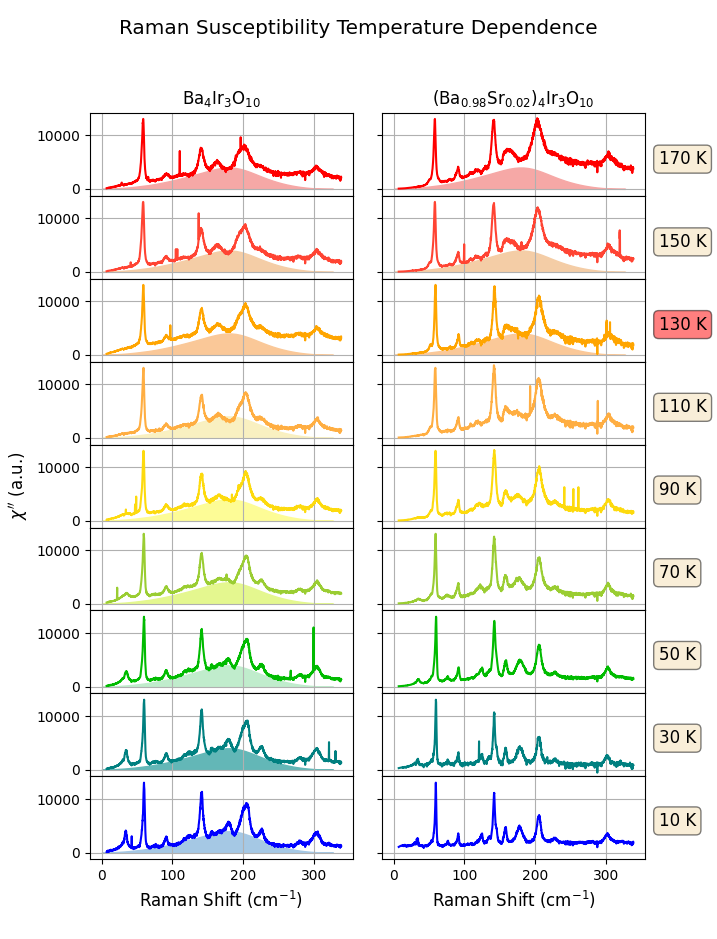}
\caption{Temperature dependence of Raman spectra in the $bb$ photon polarization for both compounds. 
 The Ba${}_4$Ir${}_3$O${}_{10}$ broadened phonons and broad 4-spinon hump at 10 K (shaded) persist up to 170 K, but are absent in the magnetically-ordered sister compound ((Ba${}_{0.98}$Sr${}_{0.02}$)${}_4$Ir${}_3$O${}_{10}$) at all temperatures below its  N{\'e}el temperature of 130 K. }
\label{fig:ramandoped}
%\end{center}
\end{figure}

Raman spectroscopy of the pure compound shows two unusual features at low temperature  (Fig.\ \ref{fig:ramantemp}b).
First, phonon peaks are extremely broad corresponding to short lifetimes. Second, a broad hump centered near 180 cm${}^{-1}$ is present underneath the phonons. The hump persists across the range of all measured temperatures, 10 K to 170 K.  Interestingly, slight Sr substitution for Ba, which induces antiferromagnetic order below 130 K, makes the phonons narrow and eliminates the broad hump (Fig.\ \ref{fig:ramantemp}c). This suggests that the hump and the reduced phonon lifetimes are associated with the spin-1/2 excitation sector. At higher temperatures above the (Ba${}_{0.98}$Sr${}_{0.02}$)${}_4$Ir${}_3$O${}_{10}$ magnetic transition $T_N$, the broad hump is restored (Fig.\ \ref{fig:ramandoped}).

Raman spectra presented here were obtained with a 671 nm solid state laser. All experiments were performed on a McPherson custom triple spectrometer equipped with a LN$_2$ cooled charge-coupled device (CCD) detector. It was configured in a subtractive mode with 1200 grooves per mm gratings in the filter stage and 1800 g/mm in the spectrometer stage. The sample was mounted in a top-loading closed-cycle refrigerator. Phonons were observed only in the spectra measured with incident and scattered photon polarization parallel to the crystallographic b-axis ($bb$ geometry) (Fig.\ \ref{fig:ramantemp}a).

\textbf{Broad hump.}
We begin by discussing the broad hump feature in the Raman signal. As evident from Fig.\ \ref{fig:ramantemp}b,c and Fig.\ \ref{fig:ramandoped}, the presence of the hump, as a function of Sr-substitution and temperature, is precisely correlated with the absence of magnetic order. This perfect correlation already strongly implies that it is associated with excitations of the non-ordered quantum liquid of spins. Indeed it is reminiscent of the broad feature reported in the Raman spectra of the 1D magnet CuGeO${}_3$, which is  characterized by 1D spin chains with spinon excitations \cite{udagawa1994, brenig1997, hase1993,van1996, braden1998, sato2012, Kiryukhin95}. 
A multi-magnon mechanism for the hump is ruled out by the absence of magnetic ordering where the hump is present. Moreover a two-magnon continuum of free magnons at zero field (particle-hole symmetry) would host a Van Hove singularity of magnon joint density of states, as has been recently observed \cite{Sala2020}, but in contrast to the featureless hump seen here. %(This divergence is remeniscient of diverging density of states in 1D systems, though here it would be expected for the two-magnon continuum regardless of spatial dimension.) 
 The observed hump necessarily arises from a fractionalized excitation, which we will refer to as a spinon, and thus represents a four-spinon continuum.

\begin{figure}[htp!]
%    \centering\noindent
    \includegraphics[width=0.5\textwidth]{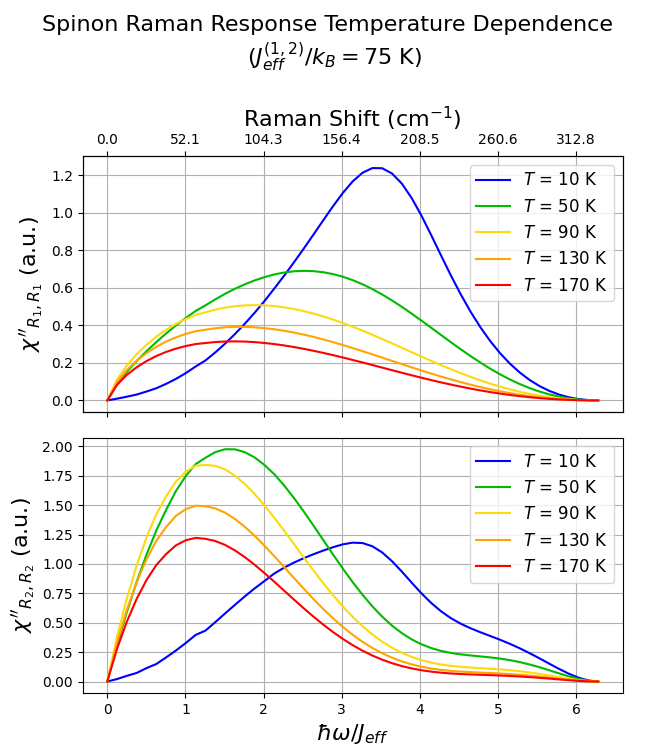}
    \caption{Spinon Raman response computed within two mean field choices $R_{\nu = 1,2}$, with $J_\text{eff}^{(\nu)}/k_B = 75$ K at $T = 10, 50, 90,130$ and 170 K (same vertical scale). The difference between $R_1$ and  $R_2$ at higher temperatures quantifies the  self-consistency breakdown of the Raman mean field theory.}
    \label{fig:mfspectrum}
%\includegraphics[scale=0.7]{figures/fig3-draft.png}
%\caption{Mean field scattering spectrum \eqref{eqn:ianal} for $R_{\nu = 1,2}$ with $J_\text{eff}/k_B = 85$ K at $k_BT/J_\text{eff} = 0.20$. A choice of $\nu$ gives uncertainty in extracting $J_\text{eff}$.}

\end{figure}

\textbf{Theoretical model via spinons of $J_1$-$J_2$-zigzag chains.} To test this hypothesis, we construct a concrete theoretical model and use it to compute the expected Raman spectra of a four-spinon continuum. The theoretical model we use entails decoupled 1D zigzag chains. Even though  Ba${}_4$Ir${}_3$O${}_{10}$ is a 2D or 3D magnet and clearly not a collection of 1D chains (neither microscopically nor phenomenologically; e.g., in its lack of Bonner-Fisher peak in susceptibility), it has been argued to be fruitfully modeled in terms of coupled 1D Heisenberg chains \cite{cao2020}, providing a tractable theoretical model for spinons. Since these spinons are 1D spinons, and not necessarily the true 2D quantum liquid spinons, we expect the resulting model to approximately capture low energy physics but fail at higher energies or temperatures. 
Raman signals from decoupled 1D Tomonaga-Luttinger liquids have been computed in bosonization \cite{sato2012}, which captures only the low-energy power-law onset of the four-spinon hump; numerically in small systems \cite{singh1996}; and in certain types of spinon mean field theories \cite{brenig1997,muthukumar1996,gomez1990}. Here we compute the four-spinon Raman response in mean field, following Ref.\ \cite{brenig1997} but also compute the mean field for an alternative equally-correct treatment which has not been done before ($R_1$ in addition to $R_2$, Fig.\ \ref{fig:mfspectrum}), and additionally point out an important restriction related to the $bb$ parallel photon polarization used here.

We thus model the spinons with an antiferromagnetic $J_1$-$J_2$ Hamiltonian \eqref{eqn:hamiltonian}:
\begin{equation}\label{eqn:hamiltonian}
    H_0 = \sum_j J_1\textbf{S}_{j} \cdot \textbf{S}_{j+1} + J_2 \textbf{S}_{j} \cdot \textbf{S}_{j+2}
\end{equation}
with $J_1,J_2 > 0$.  The quasi-1D chains of Ba$_3$Ir$_4$O$_{10}$ zigzag with a relative angle $\theta_0$ as in Fig.\ \ref{fig:ramantemp}a.
Nonzero $J_2$ and $\theta_0$ are both required in order to produce a observe a four-spinon hump for photon polarizations transverse to the chain axis.

\textbf{Result of theoretical computation.}

%\onecolumngrid
%\begin{center}
%\end{center} 
%\twocolumngrid

With this minimal model, we compute the inelastic Raman scattering spectrum using a mean field theory of free 1D fermionic spinons \cite{brenig1997} with Hamiltonian $H_0 = \sum_k \epsilon_k \hat{c}_k^\dagger \hat{c}_k$, where $\epsilon_k = -t\cos k = -\frac{\pi}{2}J_\text{eff}\cos k$. Here, we introduce an energy scale $J_\text{eff}$ which is a function of Hamiltonian parameters ($J_1,J_2$) scaled so that, in the $J_2 \to 0$ limit, it reproduces the exact Bethe ansatz result of $\epsilon_k = -\frac{\pi}{2}J_1\cos(k)$ \cite{desCloizeaux1962}. Wavevectors $k$ are in units of inverse bond length projected onto the chain axis. 

The mean field four-spinon Raman susceptibility $\chi_{R_\nu, R_\nu}''(\omega)$ is plotted in Fig.\ \ref{fig:mfspectrum}
for $\nu = 1,2$ at various temperatures. Here, $R_\nu$ is one of two equivalent choices of mean field Raman operators for the scattering spectrum, and $\chi''_{R_\nu,R_\nu}$ is the imaginary part of its associated dynamical susceptibility.

The low temperature hump feature agrees with experiment (Fig.\ \ref{fig:ramantemp}b); at higher temperatures ($k_BT/J_\text{eff} > 1$) the central frequency of the feature is lower, and the susceptibilities obtained from $R_1$ and $R_2$ become different. This difference quantifies the self-consistency breakdown of the theory at high temperatures. 

The agreement between the mean field susceptibility and experiment at low temperature allows one to extract $J_\text{eff}/k_B = 75$ K. The $J_1$-$J_2$ mean field self-consistency equation derived in Ref.\ \cite{brenig1997} relates $J_\text{eff}$ to a difference between $J_1,J_2$ in $H_0$: $J_\text{eff} \approx 1.042 J_1 - 0.8106 J_2$, and the minus sign in this expression (arising from the sublattice-rotated Jordan-Wigner transformation into spinons) allows for a given $J_\text{eff}$ to arise from substantially larger $J_1, J_2$. For example, taking $J_\text{eff}/k_B = 75$ K as a reasonable energy scale for Ba${}_4$Ir${}_3$O${}_{10}$, mean field self consistency permits $J_1/k_B = J_2/k_B = 324$ K. This scale for $J_1,J_2$ is consistent with Curie-Weiss measurements.

At higher temperatures the spinons are incoherent but nevertheless remain as the magnetic excitations. This is consistent with the broad damping feature persisting as temperature increases (Fig.\ \ref{fig:ramandoped}).  Now consider the sister magnetic material with Sr replacement. Where this magnet exhibits magnetic order, no broad peak is observed, but above $T_N$ a broad peak is seen which appears similar to the pure case at high temperature. This interpretation suggests an intriguing possibility for the sister sample: the presence of the peak above $T_N$ suggests that its magnetic transition could be considered as an instability of an incoherent parent quantum liquid state, ``existing'' (incoherently) at the high temperatures above $T_N$; strictly speaking this high temperature state is just a paramagnet, but here it evidently shows an unusually dense spectrum of strong short-ranged spin fluctuations, which could be interpreted as high-temperature relics of spinons.

As to using Raman to characterize the observed spinons, the model's agreement suggests they could be consistently modeled as 1D spinons within a low temperature effective theory, though we expect the spinons of any type of 2D quantum spin liquid phase to produce a similar four-spinon continuum hump. One intriguing possibility for a 2D quantum liquid phase with spinons closer to our model is the 2D ``Bose-Luttinger Liquid'' phase, a 2D bosonic generalization of 1D Luttinger Liquids recently introduced in Ref. \cite{lake2021}; its relevance would be resolved by observing appropriate singularities away from the Brillouin zone center.

%%%%%%%%%%%%%%%%%%%%%%%%%%%%%%%%%

\textbf{Phonon linewidth broadening through spin-phonon coupling.}

Phonon broadening due to spins that show short ranged correlation without long ranged magnetic order, combined with spin-orbit interaction \cite{liu2019}, is well known e.g.\ in Sr${}_2$IrO${}_4$ at higher temperatures \cite{Gretarsson2016}. The same effect is here seen in Sr-substituted (Ba${}_{0.98}$Sr${}_{0.02}$)${}_4$Ir${}_3$O${}_{10}$, again only at higher temperatures. Most interestingly, these broad phonons persist down to the lowest temperatures in pure Ba${}_4$Ir${}_3$O${}_{10}$ (Fig.\ \ref{fig:ramandoped}) which serves as further confirmation of the magnetic quantum liquid. This observation is very striking because increased disorder, such as introduced by Sr substitution, typically makes the phonons broader, not narrower. The low temperature narrowing of the phonon peaks upon Sr substitution, and indeed the broadening in  pure Ba${}_4$Ir${}_3$O${}_{10}$ down to the lowest temperature, provide additional evidence for a quantum liquid with spinon excitations  in  Ba${}_4$Ir${}_3$O${}_{10}$.

Beyond the particular quantum liquid state here, this behavior shows that strong coupling between magnetic degrees of freedom and phonons when the spin-orbit interaction is strong exists at the lowest temperatures, not just above $T_N$ as previously observed in Sr${}_2$IrO${}_4$.
Interestingly here the broadened phonon peaks are symmetric and do not show any significant Fano lineshape. Thus the broad hump does not have a measurable interaction with phonons; instead, it appears that phonons are broadened by magnetic fluctuations that are not Raman active. 

Phonon peak assignments can be made based on correlating the phonon energies with particular features of the crystal structure. Ba is heavy and most loosely bonded atom, so we assign the peak at low energy of 50 cm${}^{-1}$ to the Ba vibration. This peak is sharp in both samples. Substitution has a radical impact on peaks at higher energies, which correspond to various rigid motions of the IrO${}_6$ octahedra. These modes in the magnetically ordered sample narrow dramatically on cooling even though the Ba site is not affected. On the other hand, modes in the frustrated sample remain broad down to the lowest temperatures. This behavior dramatically illustrates that disorder of the pseudospins in the quantum liquid phase dominates phonon damping for a large subset of phonons. This result naturally explains coupling of the pseudospins to the phonons.

%%%%%%%%%%%%%%%%%%%%%%%%%%%%%%%%%

\textbf{Thermal conductivity and future outlook for thermoelectricity.}

The preceding discussion on phonon linewidth broadening by spinon-phonon scattering suggests that the suppressed phonon lifetimes should also be reflected in a reduced phonon contribution to thermal conductivity. Indeed the thermal conductivity shows this type of behavior, with a surprisingly small value and an increase, within a window of low temperatures, upon Sr substitution \cite{cao2020}. That this observed thermal conductivity behavior can be mostly ascribed to phonon lifetimes is shown by the weak dependence of the thermal conductivity on applied magnetic fields within the quantum spin liquid state.

Such a quantum liquid with distinct phonon damping also presents a new direction for studies of thermoelectrics, in which poor phonon thermal conductivity is essential. Attempts to control thermal conductivity are focused on engineering structures that produce flat phonon bands \cite{christensen2008avoided}. Alternatively, phonon lifetime can be reduced via damping by interactions with other phonons, defects, or electrons \cite{ikeda2019kondo}. But this type of phonon damping necessarily accompanies an unwanted consequence of reduced electrical conductivity. It is also established that spin-phonon coupling via magnetostriction can reduce the phonon lifetime, thus suppress phonon thermal conductivity \cite{Bansal2017}. Here we demonstrate that spin-orbit interaction, rather than magnetostriction, can be an efficient driver of phonon damping which extends possible thermoelectricity to a new class of materials: magnetically non-ordered 4d and 5d transition metal materials with strong spin-orbit interaction. When these exhibit strongly correlated conducting phases, rather than Mott insulators, the spin sector of the conducting electrons may still damp phonons in analogy to the spinon-phonon coupling mechanism in the present Mott insulator case.

\begin{acknowledgments}
We thank Martin Mourigal, Zhigang Jiang, and Michael Pustilnik for helpful conversations. 
I.K. acknowledges the Aspen Center for Physics where part of this work was performed, which is supported by National Science Foundation grant PHY-1607611. Work at the University of Colorado was funded by the U.S. Department of Energy, Office of Basic Energy Sciences, Office of Science, under Contract No. DE-SC0006939 and by National Science Foundation grant DMR-1903888. 
\end{acknowledgments}

%%%% END PASTE

\appendix

\section{Supplemental Material}
\subsection{
Details of theoretical computation.
}

%\iffalse
%\textbf{}
% Theory details to be moved to appendix

To derive the Raman response, we proceed as usual \cite{sato2012} by relating the inelastic Raman scattering spectrum $I(\omega)$ to the dynamical susceptibility of the relevant mean field operator $R$: $I(\omega) = \frac{1}{2\pi}\int_{-\infty}^\infty dt \ e^{i\omega t} \langle R(t) R(0) \rangle$. The scattering spectrum $I$ is related to the dynamical susceptibility $\chi''$ by the fluctuation dissipation theorem: $I(\omega) = \chi''(\omega)/(1-e^{-\omega/T})$. Here, $\omega$ is the energy of the incident or scattered photon, and the operator $R$ is the Loudon-Fleury photon-induced superexchange operator \cite{fleury1968}
\begin{equation}\label{eqn:ramanoperator}
    R = \sum_{\textbf{r}_1,\textbf{r}_2} (\mathbf{\hat{e}}_i \cdot \mathbf{\hat{r}}_{12})(\mathbf{\hat{e}}_s \cdot \mathbf{\hat{r}}_{12})A(\textbf{r}_{12}) \textbf{S}_{r_1} \cdot \textbf{S}_{r_2}
\end{equation}
For brevity, we will refer to $R$ as a Raman operator. In \eqref{eqn:ramanoperator}, $\mathbf{\hat{r}}_{1,2}$ is the unit vector pointing from lattice site $\textbf{r}_2$ to lattice site $\textbf{r}_1$, and $\mathbf{\hat{e}}_{i}$ (respectively $\mathbf{\hat{e}}_{s}$) is the polarization vector of the incoming (outgoing) photon. The factor $A(\textbf{r}_{12})$ is difficult to compute, but it is known that the ratio of $A$ on different bonds is on the order of the spin-exchange couplings on the bonds (e.g. $A(\textbf{r}_{j,j+2})/A(\textbf{r}_{j,j+1})$ is $\mathcal{O}(J_2/J_1)$) \cite{sato2012}. Given a bare Hamiltonian $H$ for a system, it is clear from the definition of $I(\omega)$ and \eqref{eqn:ramanoperator} that two Raman operators $R,R'$ yield the same spectrum if there exists some real constant $C$ such that $R' = R - CH$. If such a $C$ exists, we will say that $R$ and $R'$ are \textit{spectrally equivalent}. Having reviewed some preliminary results on inelastic Raman scattering, we may consider the particular case of an isolated 1D magnetic chain. 

Using the coordinate system of Fig.\ \ref{fig:ramantemp}, a transverse $bb$ polarization corresponds to $\theta_i = \theta_s = -\pi/2$. 
In this case
 the Raman operator is $R = R_1 \propto \sin^2\theta_0\sum_j \textbf{S}_j \cdot \textbf{S}_{j+1}$. 
We note, however, that the freedom due to spectral equivalence also allows one to choose $R = R_2 \propto \sum_j \textbf{S}_j \cdot \textbf{S}_{j+2}$. In the limit of a straight chain ($\theta_0 \to 0$), the Raman operator vanishes (up to spectral equivalence), and no spinons are seen in the spectrum. Alternatively, in the limit of $J_2 \to 0$, the Raman operator has neither the form of $R_1$ nor $R_2$. In this case, the resultant susceptibility does not exhibit a broad hump feature \cite{sato2012}. Hence, both second neighbor interactions and a zig-zagged chain are sufficient conditions to observe spinons in the scattering spectrum for photon polarizations transverse ($bb$) to the chain axis. 

Within the minimal $J_1$-$J_2$ model \eqref{eqn:hamiltonian}, we compute the inelastic scattering spectrum using a mean field theory of free 1D spinons \cite{brenig1997}. Taking $H_0 = \sum_k \epsilon_k \hat{c}_k^\dagger \hat{c}_k$ as the minimal Hamiltonian and considering $bb$ polarization, one makes a choice of the Raman operator up to spectral equivalence. That is, one chooses $R = R_\nu$ ($\nu = 1,2$). The inelastic scattering spectrum is then given by \eqref{eqn:ianal}
\begin{widetext}
\begin{equation}\label{eqn:ianal}
    I^{(\nu)}(\omega) \propto \int_{-\pi}^\pi dk \int_{-\pi}^\pi dq \ \sum_{k'} \frac{h^{(\nu)}(k,k',q)[h^{(\nu)}(k,k',q) - h^{(\nu)}(k,k',k'-k-q)]}{\sqrt{(2t\sin(q/2))^2 + (\epsilon_{k+q}-\epsilon_k-\omega)^2}} \times f(\epsilon_k)(1-f(\epsilon_{k+q})) f(\epsilon_{k'}) (1-f(\epsilon_{k'-q})) 
\end{equation}
\end{widetext}
where $f$ is the Fermi function, $h^{(1)}(k,k',q) = \cos(q)$ and $h^{(2)}(k,k',q) = \cos(2q) - \cos(2k+q) - \cos(2k'-q)$, and the sum over $k'$ is taken for all $k' \in [-\pi,\pi]$ which satisfy $2t\sin(q/2)\sin(k'-q/2) = \epsilon_{k+q}-\epsilon_k - \omega$.
%\fi

\bibliography{Ba4Ir3O10}% Produces the bibliography via BibTeX.

\end{document}